\documentclass[%
preprint,
nofootinbib,
 amsmath,amssymb,
 aps,
floatfix,
]{revtex4-1}

\usepackage{graphicx}
\usepackage{dcolumn}
\usepackage{bm}

\usepackage{url}

\newcommand{\argmax}{\operatornamewithlimits{argmax}}
\newcommand{\argmin}{\operatornamewithlimits{argmin}}

\begin{document}

\title{Efficiently detecting overlapping communities through seeding and semi-supervised learning}

\author{ Changxing Shang$^{1,2}$, Shengzhong Feng$^{1}$, Zhongying Zhao$^{1}$, and Jianping Fan$^{1}$ } 
 \email{cxshang@gmail.com,\{sz.feng,zy.zhao,jp.fan\}@siat.ac.cn}
\affiliation{
$^{1}$Shenzhen Institutes of Advanced Technology, Chinese Academy of Sciences, Shenzhen, China \\
$^{2}$University of Chinese Academy of Sciences, Beijing 100080, China \\
}


\begin{abstract}
Seeding then expanding is a commonly used scheme to discover overlapping communities in a network. Most seeding methods are either too complex to scale to large networks or too simple to select high-quality seeds, and the non-principled functions used by most expanding methods lead to poor performance when applied to diverse networks. This paper proposes a new method that transforms a network into a corpus where each edge is treated as a document, and all nodes of the network are treated as terms of the corpus. An effective seeding method is also proposed that selects seeds as a training set, then a principled expanding method based on semi-supervised learning is applied to classify edges. We compare our new algorithm with four other community detection algorithms on a wide range of synthetic and empirical networks. Experimental results show that the new algorithm can significantly improve clustering performance in most cases. Furthermore, the time complexity of the new algorithm is linear to the number of edges, and this low complexity makes the new algorithm scalable to large networks.
\end{abstract}

\pacs{Valid PACS appear here}
\maketitle


\section{\label{section:intr}Introduction} 

Many complex systems can be abstracted as networks or graphs, where the elementary parts of a system and their mutual interactions are nodes and edges (or links), respectively. A key property of many networks is their community structure: nodes with similar properties or functions have more edges than random pairs of nodes and tend to be gathered into distinct subgraphs, which are called communities (also modules or clusters). Such examples occur in many complex systems, including sociology, biology \cite{girvan2002community}, and computer science \cite{borgs2004exploring}. In reality, a node may belong to multiple communities. For example, a researcher may be active in several areas; a person usually has connections to several social groups like family, friends, colleagues, and so on. Overlapping algorithms aim to discover a cover, which is defined as a set of communities in which each node belongs to at least one community.

Local expansion and optimization is a common scheme for many methods \cite{lee2010detecting-gce,lancichinetti2009detecting-lfm} to find overlapping communities. The detecting process consists of two steps: selecting seeds and expanding the seeds to form communities. The following two paragraphs give an overview of common methods for seed selecting and expanding, respectively.

 The quality of seeds has an important impact on the final detection performance. For example, Lee et al. indicate that the performance gap between Greedy Clique Expansion (GCE) \cite{lee2010detecting-gce} and Local Fitness Maximization (LFM) \cite{lancichinetti2009detecting-lfm} is largely due to the different seed selecting method, because they use the same expanding process. When replacing a selecting method with a better one, the quality of detecting communities can be improved \cite{lee2011seeding}. Lancichinetti et al. \cite{lancichinetti2009community} also gave an example to show how seeding methods affect the expectation and maximization (EM) method proposed by Newman et al \cite{newman2007mixture}. In addition, the conclusion above is also proved by our experimental results. There are three kinds of methods that are most often used for seed selecting: random, maximal cliques, and ranking. Random methods often lead to unstable performance due to arbitrariness \cite{lancichinetti2009detecting-lfm,lee2011seeding,lancichinetti2009community}. Using maximal cliques as seeds is another commonly used method \cite{lee2010detecting-gce,shen2009detect} to get better community structures at the cost of a loss of scalability. The Bron-Kerbosch algorithm \cite{bron1973algorithm} used to find all maximal cliques is exponentially complex ($O(3^{n/3})$, where n is the number of nodes of the network). In practice, the Bron-Kerbosch algorithm may run fast on networks with nodes less than $10^5$ due to skills such as pruning technologies, but it is difficult to scale to even larger networks. Ranking methods give each node or edge a rank; a removing or appending strategy is used to select seeds. For its reputation, Pagerank is often used to compute rank values. Rank Removal (RaRe) \cite{baumes2005finding} assumes nodes with high rank do a significant amount of communication, so it sequentially removes high-rank nodes until some ``cores'' are left as seeds. The above assumption is improper because many high-rank nodes are authorities of their communities and suitable as seeds. For example, selecting node 34 in figure \ref{fig:karate} as a seed will make the community detection process easier. On the other end of the spectrum, appending methods such as Link Aggregate (LA) \cite{baumes2005efficient} select seeds in decreasing rank order. The drawback of appending methods is that many hub nodes in networks have high ranks. For example, nodes 12 and 49 in figure \ref{fig:seeding} are hubs, and expanding from them will result in poor communities. Another drawback of appending methods is that they prefer to select seeds from major communities to minor ones, so diversity of seeds cannot be guaranteed. We believe that the dilemma of ranking methods is rooted in their globally ranking behavior. This paper proposes an efficient seeding method that overcomes the drawbacks of the three kinds of methods mentioned above by first ranking edges locally and then selecting seeds globally.

Most expanding methods \cite{lee2010detecting-gce,lancichinetti2009detecting-lfm} use a local fitness function to decide whether a node should be included in a community. Yang et al. summarize 13 functions \cite{yang2012defining} based on the intuition that links in communities are dense while links between communities are sparse. An advantage of these functions is that they only use local (or neighborhood) information to decide the belonging of a node, so the expanding speed is fast. On the other hand, these functions are all heuristic and lack principled support, so they are not qualified to be used on a wide range of synthetic and empirical networks. Lee noted \cite{lee2010detecting-gce}, ``Just as there is no universally correct concept of community that spans all domains, one cannot argue that any given fitness function will be appropriate for all types of network data.'' As another drawback, each community expands independently without any negotiation with others, which may lead to highly similar communities that share a large number of nodes. Though a post-merging process can merge these communities together, the criterion of merging is difficult to decide. In this paper, a new expanding method is proposed that replaces the local fitness function with a global optimization function to infer the belonging community of an edge. Naturally, negotiations are introduced by the global function. By virtue of the wide applicability of Bayesian inference, the new method can also be applied on diverse networks.

This paper proposes an algorithm called ITEM, which uses information theory and an EM process to discover communities in a network. ITEM first transforms a network into a corpus where edges and nodes are treated as documents and terms, respectively. Then it classifies each edge into a community, and two endpoints of the edge are naturally assigned to the community. The contributions of the paper are as follows:
\begin{enumerate}
\item The concept of the Jaccard matrix of a network is proposed. Using the Jaccard matrix, the topic (i.e., the belonging community) of an edge can be extracted easily. 
\item An efficient and effective seeding method is proposed that overcomes the drawbacks of traditional methods.
\item A principled expanding method is proposed. By treating seeds selected as a training set, the semi-supervised learning technology is used to classify edges into communities.
\end{enumerate}
We conducted experiments on a wide range of synthetic and empirical networks, and the experimental results show that ITEM significantly improves clustering performance. The total computational complexity of ITEM is linear to the number of edges, which renders ITEM scalable to large networks.

The organization of the paper is as follows: section 2 introduces the skeleton of the ITEM algorithm and the most used terminologies. The seeding and expanding methods are given in sections 3 and 4, respectively. In section 5, we introduce the experimental setup, including benchmark networks and the related algorithms.  Experiment results are evaluated and analyzed in section 6. Finally, conclusions and suggestions for future research are provided. To reproduce the results, we published ITEM's code on the web \footnote{\url{https://github.com/cxshang/ITEM}}. 

\section{Terminology and skeleton of ITEM Algorithm} \label{sec:skeleton}

\begin{figure}[!ht]
  \centering
   \includegraphics[width=75mm,height=50mm]{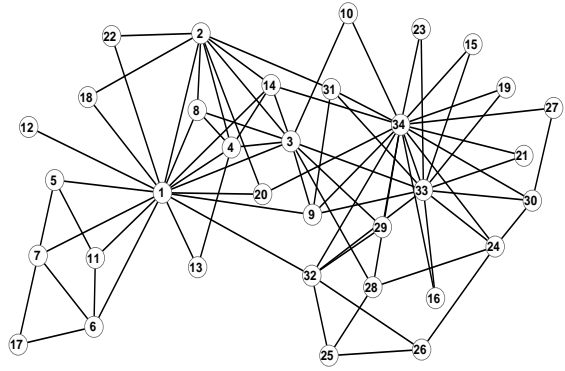} 
 \caption{The Karate network, which represents friendship between members of a university sports club. The club is divided into two communities as a result of an internal dispute.}
\label{fig:karate}
\end{figure}

In this section, some terminology used in the paper is given in table \ref{tab:terms}. There are two notations to denote an edge, so when its two endpoints (e.g., $v_i$ and $v_j$) must be explicitly given, double subscript notation (e.g., $e_{i-j}$) is used; otherwise, we prefer single subscript notation (e.g., $e_i$ denotes the $i$th edge). To clearly illustrate these terminologies, the Karate network in figure \ref{fig:karate} is used as an example. The skeleton of ITEM is also explained to help readers rapidly understand the main idea of ITEM.

\begin{table}[!ht]
\centering
\begin{tabular}{|p{3.5cm}|p{10cm}|p{3.5cm}|}
\hline
Terminology&Definitions&Examples  \\ 
\hline 
Graph or Network & A graph or a network is an ordered pair $G = (V, E)$ comprising a set $V$ of vertices or nodes together with a set $E$ of edges or links. & Karate network in figure \ref{fig:karate}.  \\
\hline
$m$ &$m=|E|$ is the number of edges. &78  \\
\hline
$n$ &$n=|V|$ is the number of nodes. & 34 \\
\hline
$v_i$ &$v_i$ represents the $i$th vertex of $G$. Where $i \in \{1,2,\cdots,n\}$. & $v_1 \newline v_{34}$\\
\hline
$e_i$(single subscript notation) &$e_i$ represents the $i$th edge of $G$. Where $i \in \{1,2,\cdots,m\}$. & $e_1 \newline e_{78}$ \\
\hline
$e_{i-j}$(double subscript notation) &$e_{i-j}$ represents the edge which has two endpoints $v_i$ and $v_j$. & $e_{1-4} \newline e_{33-34}$\\ 
\hline
$Nb(v_i)$ &$Nb(v_i)$ is a set of nodes containing all neighbor nodes of $v_i$. & $Nb(v_7)=\{v_1, v_5, \newline v_6,v_{17} \}$ \\
\hline
 $Nb(e_{i-j})$ or $Nb(e_k)$, suppose two endpoints of $e_k$ are $v_i$ and $v_j$. &$Nb(e_k)=Nb(e_{i-j})=(Nb(v_i) \cap Nb(v_j)) \cup \{v_i, v_j\}$ these are the neighbor nodes of $e_{i-j}$ or $e_k$. & $Nb(e_{12})=Nb(e_{6-7})=\{v_1,v_6,\newline v_7,v_{17} \}$, suppose $e_{6-7}$ is the $12$th edge. \\ 
\hline
 $Ic(v_i)$ &  $Ic(v_i)$ denotes the incident edges of $v_i$. &  $Ic(v_7)=\{e_{1-7}, e_{5-7}, \newline e_{6-7}, e_{7-17} \}$ \\
\hline
$Ic(e_{i-j})$ or $Ic(e_{k})$, suppose two endpoints of $e_k$ are $v_i$ and $v_j$. &The incident edges of $e_{k}$ or $e_{i-j}$ are defined as $Ic(e_{k})=Ic(e_{i-j})=Ic(v_i)\cup Ic(v_j)-\{e_{ij}\}$. &$Ic(e_{12})=Ic(e_{6-7})=\{e_{1-6},e_{1-7}, \newline e_{5-7}, e_{6-11}, \newline e_{6-17}, e_{7-17} \}$ suppose $e_{6-7}$ is the $12$th edge. \\
\hline
 $deg(v_i)$ &$deg(v_i)$ represents the degree of $v_i$. &  $deg(v_7)=4$ \\
\hline
\end{tabular}
\caption{Terminologies used in the paper.}
\label{tab:terms}
\end{table}
ITEM exploits text-mining technologies \cite{berry2004survey} to discover the communities of a network. We propose the concept of the Jaccard matrix of a network by observing that the community of an edge can be largely determined by its two endpoints and their shared neighbors, just as the topic of a document can be identified by a few key terms or features \cite{koller1997hierarchically}. The origin of the Jaccard matrix is motivated by the Jaccard index, which is a statistic commonly used to measure the similarity of two endpoints of an edge. For the similarity between two endpoints of $e_{i-j}$, the Jaccard index is defined as $J(v_i, v_j)=\frac{Nb(v_i) \cap Nb(v_j)}{Nb(v_i) \cup Nb(v_j)} $.

The Jaccard matrix of a network $G$ is denoted as $M$ and is an $m \times n$ matrix (for meanings of $m$ and $n$, please see table \ref{tab:terms}). For each item $M_{ij}$ where $i \in \{1,2,\cdots m\}$ and $j \in \{1,2,\cdots n\}$,
\begin{equation}
M_{ij} = \left\{
\begin{array}{l l}
0 & \quad \text{if $v_j \notin Nb(e_i)$ }\\
w_{j} & \quad \text{if $v_j \in Nb(e_i)$ }\\ 
\end{array} \right.
\end{equation}
$w_{j}$ can simply be assigned a value of 1 or using the $tf\cdot idf$ \cite{robertson2004understanding} method. Table \ref{tab:jaccard} displays a fraction of entries of Karate's Jaccard matrix.

The following lists the features of the Jaccard matrix. First, a document $e_{i-j}$ only includes terms in $Nb(e_{i-j})$; all other neighbors of $v_i$ and $v_j$ are discarded. The discarding operation resembles the preprocessing step in text mining that removes low-frequency words from a document. After the discarding operation, the topic of document $e_{i-j}$ is easier to identify. The clarity comes from the fact that edges are more specific to a certain community or topic than nodes. Second, because of the increased density of a community and the more common neighbors shared by its edges, this produces many similar documents in the Jaccard matrix. Based on the similarity, we can cluster the edges in the same community together.

\begin{table}[!ht]
\centering
\begin{tabular}{|c|p{0.35cm}|p{0.35cm}|p{0.35cm}|p{0.35cm}|p{0.35cm}|p{0.35cm}|p{0.35cm}|p{0.35cm}|p{0.35cm}|p{0.35cm}|p{0.35cm}|p{0.35cm}|p{0.35cm}|}
\hline 
\  &1 &2 &3 &4 &8 &14 &18 &20 &24 &28 &30 &33 &34 \\
\hline 
$e_{2-3}$ & 1 &1 &1 &1 &1 &1 & & & & & & & \\
\hline 
$e_{2-4}$ & 1 &1 &1 &1 &1 &1 & & & & & & & \\
\hline 
$e_{2-8}$ & 1 &1 &1 &1 &1 & & & & & & & & \\
\hline 
$e_{2-14}$ & 1 &1 &1 &1 & &1 & & & & & & & \\
\hline 
$e_{1-18}$ & 1 &1 & & & & &1 & & & & & & \\
\hline 
$e_{20-34}$ & & & & & & & &1 & & & & &1 \\
\hline 
$e_{24-28}$ & & & & & & & & &1 &1 & & &1 \\
\hline 
$e_{24-33}$ & & & & & & & & &1 & &1 &1 &1 \\
\hline 
$e_{24-34}$ & & & & & & & & &1 &1 &1 &1 &1 \\
\hline 
$e_{24-30}$ & & & & & & & & &1 & &1 &1 &1 \\
\hline
\end{tabular}
\caption{A fraction of entries of the Jaccard matrix of the Karate network. Each row or column corresponds to an edge or a node, respectively. All $w_j$ values are set to 1 for clarity. A blank cell in the 2nd row and 8th column indicates $v_{18}$ is not included in $Nb(e_{2-3})$, so its value is set to 0 and not displayed, and so on.}
\label{tab:jaccard}
\end{table}

 ITEM resorts to semi-supervised learning technology to classify edges into different communities. Many machine learning researchers have found that semi-supervised learning can considerably improve learning accuracy because it exploits both labeled and unlabeled information \cite{zhu2006semi}. As for the classifier, NB (Na\"{\i}ve Bayes) is used for its simplicity and effectiveness in text classification \cite{domingos1997optimality}. ITEM first selects some seeds as a training set, then an EM process is used to expand edges into communities. In each expectation step, some edges previously unlabeled get new labels (i.e., communities), and some edges change their belonging communities. Because the above unlabeled edges are then poured into the maximization step to model or refine the NB classifier, in this sense, ITEM uses semi-supervised learning to expand communities. The EM iterations are stopped until a predefined condition is matched.

\begin{figure*}[!ht]
  \centering
  \begin{tabular}{ccc}
   \includegraphics[width=55mm,height=50mm]{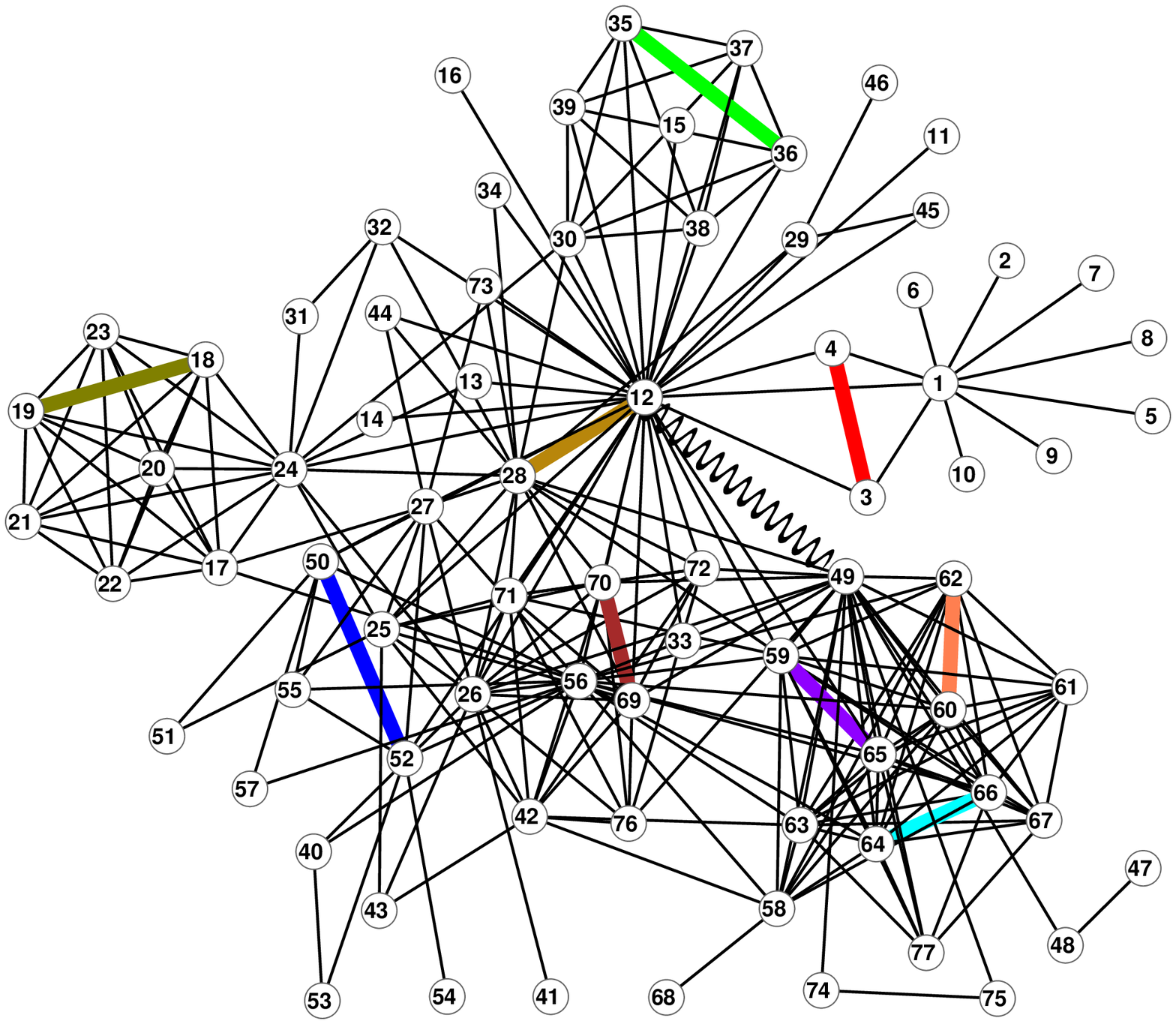} & 
   \includegraphics[width=55mm,height=50mm]{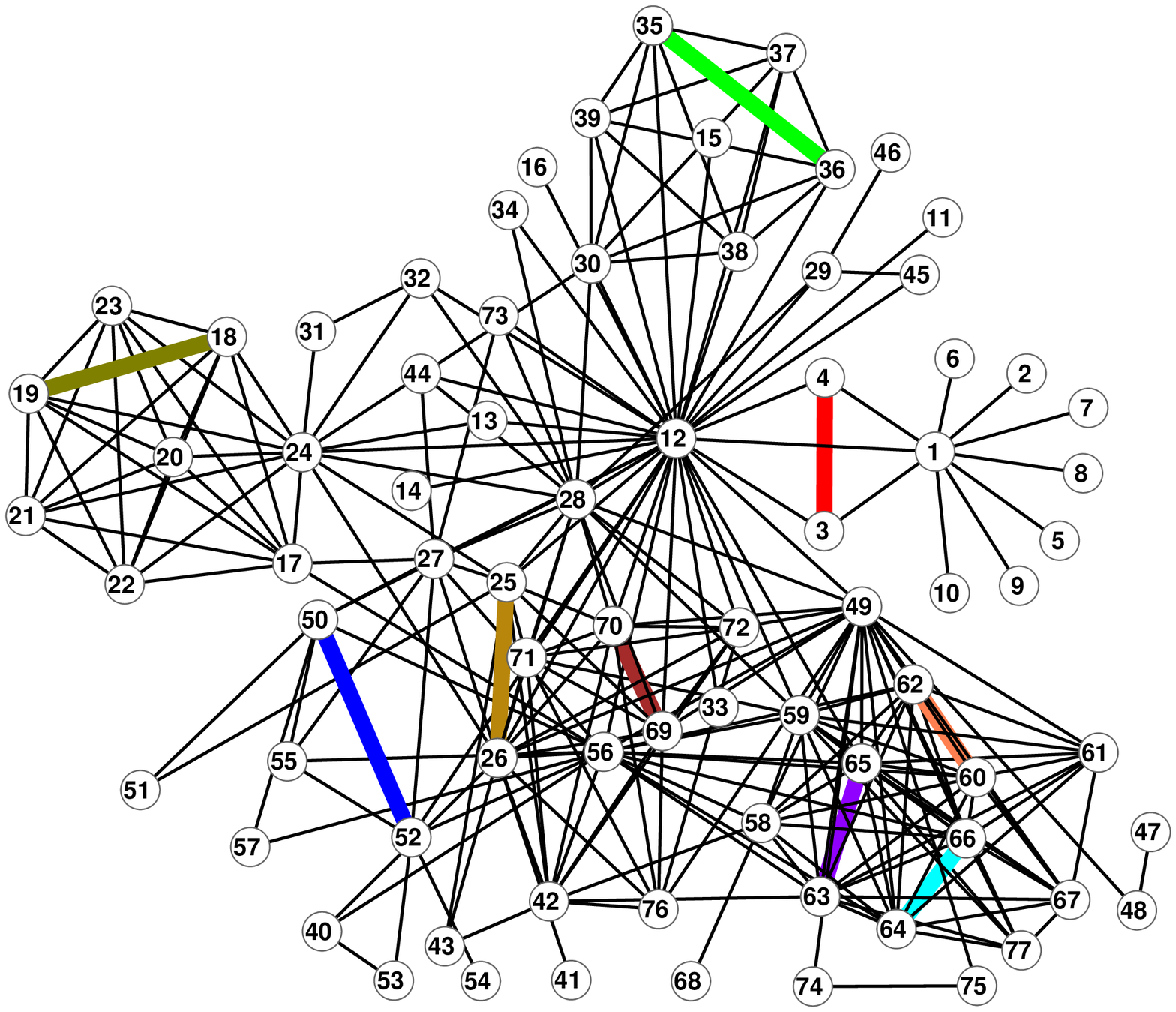} &
   \includegraphics[width=55mm,height=50mm]{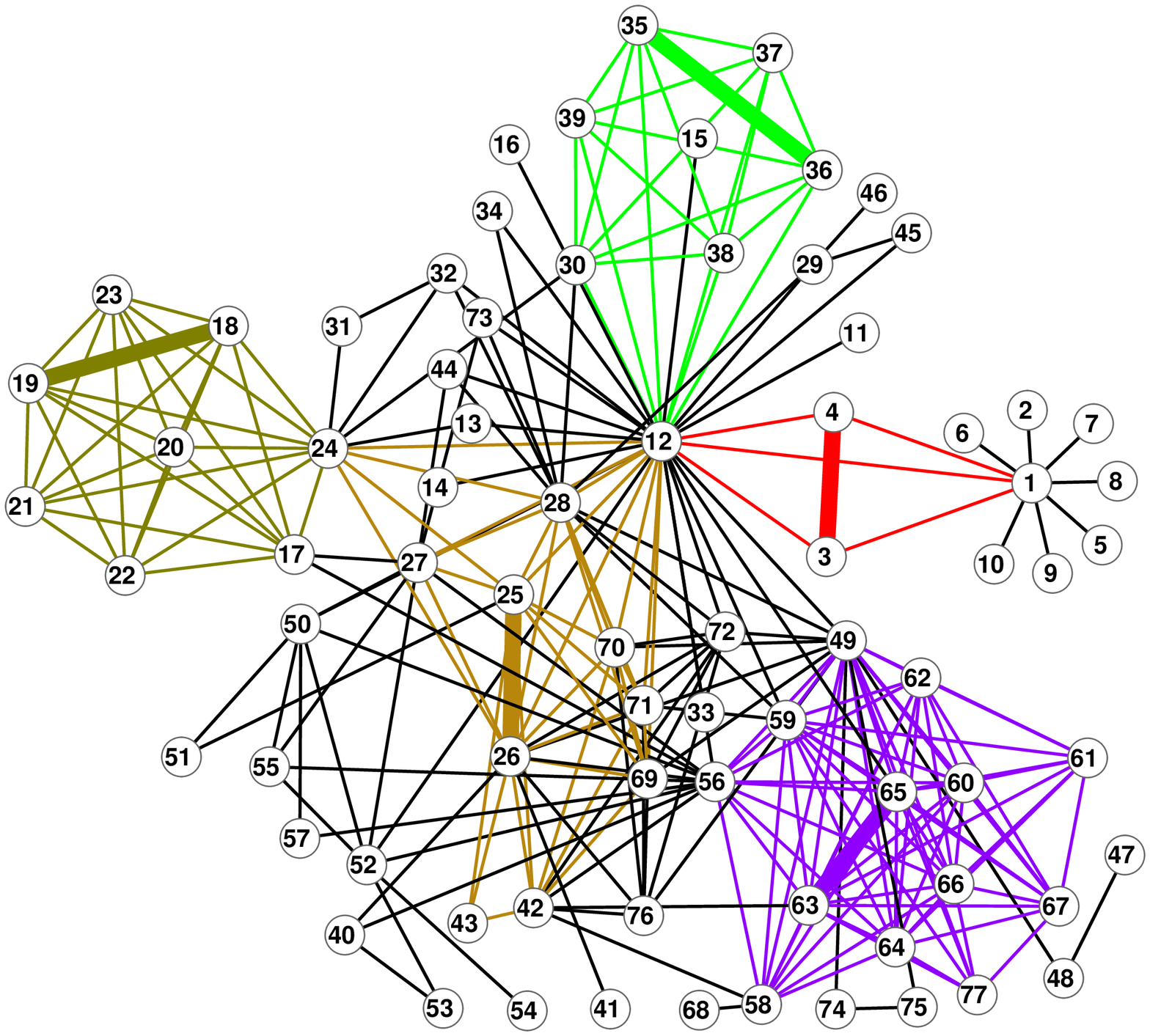} 
 \end{tabular}
  \caption{This figure illustrates the steps of the seeding process by using the LM network as an example. The LM network is a co-appearance network of characters in the novel \textit{Les Mis\'erables} by Victor Hugo. In the left subfigure, $e_{12-49}$ (the sinewave edge) is the seed selected by only using reputation scores. The color edges are the seeds selected by using reputation$\times$strength scores. In the middle subfigure,  the color edges are the seeds selected by using reputation$\times$strength$\times$specificity scores. In the right subfigure, the bold color edges are the seeds selected by using the MGIG method; the thin color edges are added later to enlarge the training examples.}
\label{fig:seeding}
\end{figure*}

\section{The seeding process} \label{section:seed}

ITEM uses two steps to select seeds. In the first step, to avoid the high computational complexity and the drawbacks of global ranking methods, a local ranking method called RSS (Reputation, Strength, Specificity) is proposed. ``Local'' has two meanings: first, RSS gives each edge a score by only using its local information (i.e., the neighbors' information). Second, edges compare their RSS scores locally. More specifically, $e_i$ only compares its score with its incident edges (i.e., $Ic(e_i)$). If there exists any one incident edge whose score is higher than $e_i$, then $e_i$ is filtered out and cannot be selected as a seed. If two edges have equal scores, then the edge with the small index is left, and the large one is filtered out. In the second step, the candidate seeds selected by RSS are fed to the maximizing global information gain (MGIG) method \cite{shang2013feature}, which is used to select distinctive and representative seeds from candidates from a global view.
  
\subsection{Selecting candidate seeds with the RSS method}

The LM network in figure \ref{fig:seeding} is a bit more complicated than Karate. In this subsection, we mainly use it is to explain the motivations of the RSS method. 

When selecting seeding nodes, completely using or ignoring reputation values is improper because both hub nodes and specificity nodes may have high reputations. Reputation and specificity also make sense for edges, and edges have another property that is lacking in nodes. We define strength to measure the link intensity between two endpoints of an edge. Intuitively, $e_{12-49}$ in figure \ref{fig:seeding} is not suitable as a seed because it has a high reputation but low specificity. $e_{63-65}$ is more suitable as a seed than $e_{74-75}$ because $e_{63-65}$ is more reputable and stronger than $e_{74-75}$. But what causes the above intuition to emerge? In fact, the number of common neighbors between two endpoints of an edge is a good indicator to measure the extent to which the edge is suitable as a seed. In the following, the formal definitions of reputation, strength, and specificity are given, all of which exploit the common neighbors concept directly or indirectly. The seeds selected by using reputation, strength, and specificity scores are listed in figure \ref{fig:seeding}.
 
To get the RSS score of each edge, the similarity of two incident edges must be calculated in advance. SimHash \cite{charikar2002similarity} is a common technology used to evaluate the similarity between the two documents. This paper uses the SimHash technology to convert document $e_i$ to a 64-bit binary number, which is denoted as $fp_i$ and called the fingerprint of $e_i$. We do not explain SimHash in detail due to space constraints. For more information, please see \cite{charikar2002similarity,manku2007detecting}. 

 For each $e_i$, where $ i\in \{1,2,\cdots,m \}$, its reputation score is defined as follows:
\begin{equation}
reputation(e_i)=\frac{1}{64}\sum_{\{j|if\ e_j \in Ic(e_i)\}}{64-hd(fp_i,fp_j)} 
\label{reputation}
\end{equation}
where $fp_i$ and $fp_j$ are the fingerprints of $e_i$ and $e_j$ respectively and  $hd(fp_i,fp_j)$ evaluates the hamming distance between $fp_i$ and $fp_j$. Equation (\ref{reputation}) gives a high reputation score to $e_i$ if a large number of nodes are shared between $Nb(e_i)$ and $Nb(e_j)$ (which makes $hd(fp_i,fp_j)$ small) or if $|Ic(e_i)|$ is large. 

The strength for $e_{i-j}$ is defined as follows: 
\begin{equation}
strength(e_{i-j})=\frac{|Nb(e_{i-j})|-1}{max(deg(v_i),deg(v_j))}
\label{strength}
\end{equation} 
 The $strength(e_{i-j})$ measures the intensity of the connection between $v_i$ and $v_j$. 

For each node, its specificity score is defined as follows:
\begin{equation*}
specificity(v_i)=\frac{\sum_{\{j|if \ e_j\in Ic(v_i)\}}{strength(e_j)}}{|Ic(v_i)|} 
\label{nodespecificity}
\end{equation*}
$specificity(v_i)$ measures the average similarity between $v_i$ and its neighbors. The following gives the specificity definition of an edge:
\begin{equation*}
specificity(e_{i-j}) 
=min(specificity(v_i), specificity(v_j)) .
\label{edgespecificity}
\end{equation*}

For $\forall e_i \in E$, its reputation, strength, and specificity scores are all in $(0,1]$. Now, the RSS score of $e_i$ can be achieved:
\begin{equation*}
\begin{split}
RSS(e_{i}) 
=reputation(e_{i}) \times strength(e_{i}) \times specificity(e_{i}) .
\end{split}
\end{equation*}

The computational complexity of RSS is $O(dm)$, where $d$ is the average degree of nodes. The low computational complexity mainly contributes to the locality of RSS. On the other hand, only comparing with its incident edges makes an edge easy to select as a seed. As a result, some similar edges appear. For example, $e_{60-62}$, $e_{63-65}$, and $e_{64-66}$ are similar, so using them as seeds may split an integral community apart. As another side effect, an edge may be thrashed among its adjacent communities, which will slow the convergence process of the subsequent EM process. To overcome this drawback, similar edges should be filtered out. In section \ref{mgigsection}, how to select final edges is explained in detail. In the following, we call the edges selected by RSS candidate seeds because the final seeds are selected from them.

\subsection{Selecting final seeds with the MGIG method} \label{mgigsection}

 MGIG is an efficient feature selection method for text classification that effectively selects distinctive and representative terms \cite{shang2013feature}. Due to the duality between terms and documents \cite{dhillon2003information}, MGIG is used to select dissimilar edges (documents) in the paper. The idea behind MGIG is very simple. In table \ref{tab:jaccard}, the document representations of $e_{2-3}$ and $e_{2-4}$ are identical, so the information lost when merging them as a virtual document (i.e., the document that includes all terms in  $e_{2-3}$ and $e_{2-4}$) is 0. Splitting the virtual document apart also releases no information. On the other hand, there is more information released when splitting $e_{2-3}$ and $e_{24-30}$ apart because they are dissimilar. MGIG tries to select edges that release the maximum information. 

MGIG selects seeds one by one. Suppose $l$ is the number of candidate seeds. If $k-1$ ($ 1 \leq k-1 < l$) edges have been selected, then MGIG selects the $k$th edge from unselected candidates, which creates the following maximum:
\begin{equation}
 p(\tilde{e}_{\mathbb{S}_{k}}) H(\mathbf{p}(V|\tilde{e}_{\mathbb{S}_{k}})) 
- p(e_{k}^*) H(\mathbf{p}(V|e_{k}^*)) \ ,
\label{selectiontj}
\end{equation}
Noting $e_k^*$ represents the $k$th edge selected from $l$ candidates. Where $H(\mathbf{p}(V|e_{k}^*))$ is the entropy of $\mathbf{p}(V|e_{k}^*)$, and $\mathbf{p}(V|e_i^*)=(p(v_1|e_i^*),p(v_2|e_i^*),\cdots,p(v_n|e_i^*))$ is the node distribution for a given edge, $e_i^*$. $\mathbb{S}_k=\left\{e_1^*,e_2^*,\cdots,e_k^*  \right\}$ is the set of candidates already selected, and $\tilde{e}_{\mathbb{S}_k}$ is just a virtual edge if we view all $k$ selected edges in $\mathbb{S}_k$ as a whole. Hence,
\begin{equation*}
 p(\tilde{e}_{\mathbb{S}_k}) = \sum_{i=1}^k p(e_i^*) \ ,
\label{eq:ibweight}
\end{equation*}
and
\begin{equation*}
 p(v_j|\tilde{e}_{\mathbb{S}_k}) = \sum_{i=1}^k  \frac{p(e_i^*)}{p(\tilde{e}_{\mathbb{S}_k})} p(v_j|e_i^*)  \  .
\label{eq:ibconprob}
\end{equation*}
For $p(v_j|e_i^*)$, it is as follows:
\begin{equation*}
p(v_j|e_i^*) = \left\{
\begin{array}{l l}
0 & \quad \text{if $v_j \notin Nb(e_i^*)$ }\\
1/|Nb(e_i^*)| & \quad \text{if $v_j \in Nb(e_i^*)$ }\\ 
\end{array} \right.
\end{equation*}
Because criterion (\ref{selectiontj}) does not apply to $k=1$, we select the first edge which makes $ H(\mathbf{p}(V|e_{1}^*))$ maximum. 

After $e_k^*$ is selected, each unselected candidate edge $e^u$ is checked. $Nb(e^u)-Nb(e_k^*)$ is the set of nodes that are included in $Nb(e^u)$ but not included in $Nb(e_k^*)$. Clearly, the greater $|Nb(e^u)-Nb(e_k^*)|/|Nb(e^u)|$, the more likely $e^u$ is to qualify as a seed. We introduce a scaling parameter $th \in \{0.1, 0.2, 0.3, 0.4, 0.5, 0.6, 0.7, 0.8, 0.9, 1.0\}$ to control whether $e^u$ should be filtered out. If $|Nb(e^u)-Nb(e_k^*)|/|Nb(e^u)| < th$ is met, $e^u$ is filtered out. There is no universal value for $th$ because of the diversity of networks and multiple scales of community structure for a network, but for a specific network, the effective range of $th$ is typically narrow.
 
In the right subfigure of figure \ref{fig:seeding}, the bold color edges are the final seeds selected by using the MGIG method with $th=0.3$. Now we get a training set that has 5 seeds, but it is too few because it contains only one training example for each community. To make the subsequent semi-supervised learning process more reliable, the training set is enlarged as follows: suppose $e^*$ is a final seeding edge, for $\forall v_i, v_j \in Nb(e^*)$; if $\exists e_{i-j} \in E$, then $e_{i-j}$ is added in the training set and labeled with a community of $e^*$. In the same subfigure, these added edges are displayed in thin color lines. We call edges of the same color a committee. In the expanding process, committees are expanded to form communities. In the following, $K$ is denoted as the final number of communities. For the LM network in figure \ref{fig:seeding}, $K=5$. 

 Suppose $p=\frac{1}{m}\sum_{i=1}^m|Nb(e_i)|$; the average number of neighbors for an edge. To select $K$ seeds, the computational complexity of using MGIG is $O(plK)$. If taking the RSS into account, the total computational complexity of the seeding process is $O(dm+plK)$, where $p \leq d+1$ and $l \ll m$. Clearly, our seeding method is more to scale than the method of finding all maximal cliques.

\section{The Expanding process} \label{section:expand}

The expanding process resorts to an EM algorithm to classify edges into communities. Details about the expectation and maximization steps are given below.

\subsection{The expectation step} \label{section:exp}

The expectation step exploits both topological and topic information to decide the most suitable community for $e_i$. We first use the topological information to judge whether $e_i$ is a potential edge of $C_k$. The potential edges of a community $C_k$ are the edges that are not included in $C_k$ currently but that may be added into $C_k$ in the expectation step. If $e_i$ is a potential edge of $C_k$, then the topic information is used to evaluate the posterior probability $p(C_k|e_i)$. Lastly, $e_i$ is assigned to the community that makes $p(C_k|e_i)$ the maximum. The community is denoted as $C^*(e_i)$.

 Suppose $Nd(C_k)=\cup_{e_{i-j}\in C_k}\{v_i,v_j\}$ is the set of nodes included in $C_k$ at present. Clearly, if neither $v_i$ nor $v_j$ is included in $Nd(C_k)$, then $e_{i-j}$ isn't adjacent to $C_k$ and cannot be a potential edge. On the other hand, if both $v_i$ and $v_j$ are included in $Nd(C_k)$ but $e_{i-j}$ isn't included in $C_k$, $e_{i-j}$ is destined to be a potential edge of $C_k$. As the third status, $e_{i-j}$ may sit between the two extremes. That is to say, $Nd(C_k)$ only includes one endpoint of $e_{i-j}$. Without loss of generality, we assume that $v_i$ is included in $Nd(C_k)$ but $v_j$ isn't. In this case, $e_{i-j}$ is treated as a potential edge of $C_k$ if it matches any condition below: 
\begin{enumerate}
\item $od(e_{i-j})>0$ and $|Nb(e_{i-j}) \cap Nd(C_k)| \geq 2$,
\item $od(e_{i-j})=0$ and $od(v_j)=0$ and $ |Nb(e_{i-j}) \cap Nd(C_k)| \geq  \frac{1}{2} deg(v_j) $,
\end{enumerate}
where $od(e_{i-j})$ is the order of $e_{i-j}$ and is defined as $od(e_{i-j})=|Nb(e_{i-j})|-2$, which is the number of triangles that include $e_{i-j}$ as one edge. As examples in figure \ref{fig:karate}, $od(e_{20-34})=0$, $od(e_{1-18})=1$, and $od(e_{2-3})=4$. $od(v_j)=\frac{1}{2}\sum_{e_i \in Ic(v_j)}od(e_i)$ is the order of $v_j$, which is the number of triangles that use $v_j$ as a vertex, so 0 order nodes are not included in any triangles.

Conditions 1 and 2 above give the constraints for non-zero order edges and zero order edges, respectively. Condition 1 indicates that at least 2 sponsors want to pull $v_j$ into $C_k$. In addition, these 2 sponsors and $v_j$ are the three vertexes of a triangle. Condition 2 imposes a rigorous requirement on a zero-order edge. If $od(e_{i-j})=0$ but $od(v_j)>0$, it means $v_j$ has other more intimate neighbors than $v_i$, so $v_j$ more likely will be pulled into a community (not necessarily $C_k$) by other links but not by $e_{i-j}$. If $od(e_{i-j})=0$ and $od(v_j)=0$, $v_j$ treats its neighbors equally, then $|Nb(e_{i-j}) \cap Nd(C_k)| \geq  \frac{1}{2} deg(v_j) $ ensures that only for at least half the neighbors of $v_j$ included in $C_k$ can $e_{i-j}$ be viewed as a potential edge of $C_k$. 

After ensuring $e_i$ is a potential edge of $C_k$, the posterior probability $p(C_k|e_i)$ is evaluated by using an NB classifier, and the final belonging community of $e_i$ is as follows:
\begin{equation}
  \begin{split}
&C^*(e_i)=\argmax_{C_k} {p(C_k|e_i)} \\
=&\argmax_{C_k} p(C_k) \underset{v_j \in Nb(e_i)}\Pi p(v_j|C_k)^{w_j} \ ,
\end{split}
\label{eq:probedge}
\end{equation}
where $w_j$ is the weight of $v_j$; its value is set using the $tf \cdot idf$ method. $p(C_k)$ and $p(v_j|C_k)$ will be evaluated in the maximization step. 

Suppose $W_{e_i}= \sum_{ \{j|if \  v_j \in Nb(e_i) \}} w_j $; then (\ref{eq:probedge}) can be written in the following form by taking logarithms, dividing by $W_{e_i}$, and adding $H(\mathbf{p}(V|e_{i}))$:
\begin{equation}
\begin{split}
&C^*(e_i) \\
=&\argmin_{C_k} KL(\mathbf{p}(V|e_i), \mathbf{p}(V|C_k))-\frac{1}{W_{e_i}}log(p(C_k))
\end{split}
\label{eq:probedgekl}
\end{equation}
where $ KL(\mathbf{p}(V|e_i), \mathbf{p}(V|C_k))$ is the Kullback-Leibler divergence between $ \mathbf{p}(V|e_i)$ and $\mathbf{p}(V|C_k)$. Hence, our expanding method resembles the K-means algorithm by using (\ref{eq:probedgekl}) to measure distances and tries to minimize a global objective function as follows:
\begin{equation}
\sum_{e_i \in PS}  KL(\mathbf{p}(V|e_i), \mathbf{p}(V|C^*(e_i)))-\frac{1}{W_{e_i}}log(p(C^*(e_i))) \ .
\label{kmeansfunc}
\end{equation}
Where $PS$ denotes the set of potential edges.


\subsection{The maximization step} \label{section:max}
In the maximization step, the unknown parameters are evaluated based on all the currently labeled edges. $p(C_k)$ is just the proportion of edges in $C_k$ versus edges in all current communities. $p(v_j|C_k)$ is evaluated as follows:
\begin{equation}
p(v_j|C_k)=\frac{w_{jk}}{\underset{\{t|if\ v_t \in Nd^{*}(C_k)\}} \sum w_{tk}} \ .
\label{eq:probclass}
\end{equation}
In the above equation, $Nd^{*}(C_k)$ is indeed the set of terms occurring in the documents of $C_k$, which is defined as $Nd^{*}(C_k)=\cup_{e_i \in C_k} Nb(e_i)$. Clearly, $Nd(C_k) \subseteq  Nd^{*}(C_k)$.

$w_{tk}$ is the weight of $v_t$ in $C_k$. A simple way is to set $w_{tk}$ to $pr(v_t|C_k)$, where $pr(v_t|C_k)$ is the conditional probability of $v_t$ occurring given $C_k$. For example, in the right subfigure of figure \ref{fig:seeding}, $pr(v_{12}|C_{red})=6/24$ ($C_{red}$ indicates the red community) because term $v_{12}$ occurs in 6 documents, and there are 24 total occurrences for terms $v_{1}, v_{3}, v_{4}$, and $v_{12}$. Unfortunately, directly using $pr(v_t|C_k)$ has a drawback because hub nodes may have higher weights than more specific nodes. For example, for the community of a dark goldenrod color in the same subfigure, the weight of $v_{12}$ will be higher than $v_{28}$'s. To bias the weights to the high-specificity nodes, a probability ratio \cite{mladenic1999feature} is introduced to assign a value to $w_{tk}$:
\begin{equation*}
w_{tk}=\frac{pr(v_t|C_k)}{pr(v_t|\overline{C_k})} 
\end{equation*}
Above, $\overline{C_k}$ represents all other communities except $C_k$. $pr(v_t|C_k)$ and $pr(v_t|\overline{C_k})$ are the conditional probabilities of $v_t$ occurring given the ``positive'' community $C_k$ and the ``negative'' community $\overline{C_k}$, respectively. 


\section{Experimental setup} 
This section consists of 2 parts. We explain the synthetic and empirical networks and performance measures in subsection 1. In subsection 2, we give a brief introduction of ITEM's comparing algorithms.

\subsection{The synthetic and empirical networks} \label{sectionsetupnetwork}
\begin{figure*}[!ht]
  \centering
\includegraphics[width=175mm,height=200mm]{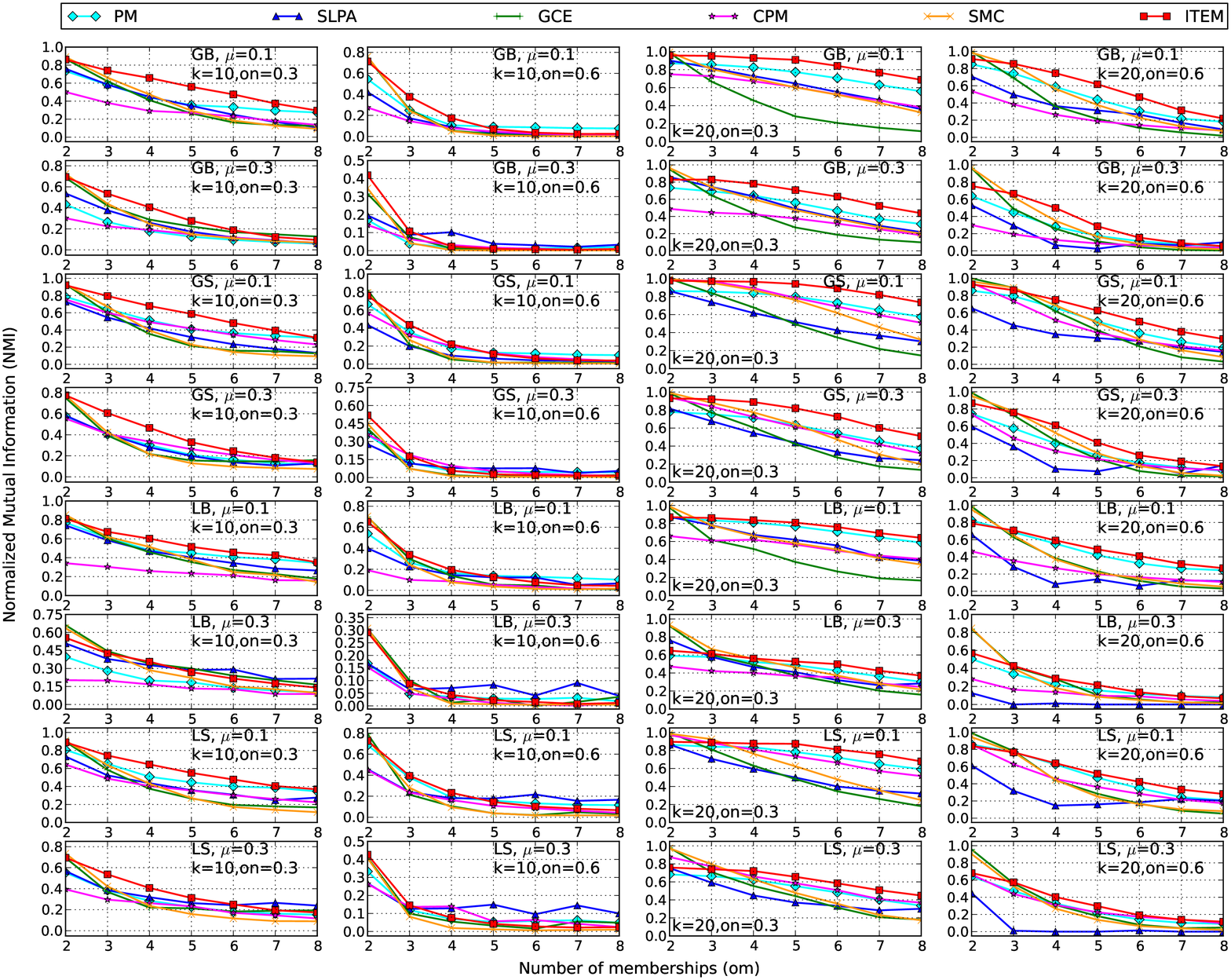}
 \caption{NMI comparisons when $om$ increases.}
\label{fig:om}
\end{figure*}

Synthetic networks are commonly used to evaluate the performance of community detection algorithms because the ground truth communities are clear. We use the Lancichinetti-Fortunato-Radicchi (LFR) benchmark \cite{lancichinetti2008benchmarks, lancichinetti2009benchmarks} to construct synthetic networks, which provides a rich set of parameters to control the network topology. Throughout the experiments, the maximum degree of LFR networks is set at 50; node degrees and community sizes are governed by power law distributions with exponents $\tau_1=2$ and $\tau_2=1$. The network sizes can be 5000 or 1000, which are denoted as G (great) or L (little) communities, respectively. The community sizes of a network can vary in small range [10,50] or big range [20, 100], denoted as S and B, respectively. For other parameters such as the mixing parameter $\mu$, which is the expected ratio between the number of boundary edges (edges whose two endpoints aren't in same community) and the number of incident edges for each node, the average degree $k$, the number of memberships of the overlapping nodes $om$, and the fraction of overlapping nodes $on$; their values are explicitly given in figure \ref{fig:om}. For each parameter setting above, we generated 10 instances. Hence, the results in figures \ref{fig:om} are the averages over 10 LFR networks. To measure the similarity of ground truth communities and found communities, normalized mutual information (NMI) \cite{lancichinetti2009detecting-lfm} is used. The NMI value is between 0 and 1, with 1 corresponding to a perfect match between the true community and the found community.

For empirical networks, we use the Facebook100 dataset \cite{traud2011comparing,traud2012social}, which includes 100 friendship networks. These networks have meta-data such as year and dorm for nodes. Evaluating these networks is difficult because the ground truth communities are unknown, so NMI cannot be used on the Facebook100 dataset. Recently, Lee and Cunningham proposed a machine learning framework \cite{lee2014community} to measure how well communities were detected based on the assumption that ``if a community detection algorithm is functioning well, then a classifier should be able to use the set of detected communities to infer missing values of a node attribute that is closely related to community structure.'' So when a classifier gives a higher accuracy score, it indicates the found communities are better.

\begin{figure*}[!ht]
  \centering
  \begin{tabular}{cc}
\includegraphics[width=105mm,height=50mm]{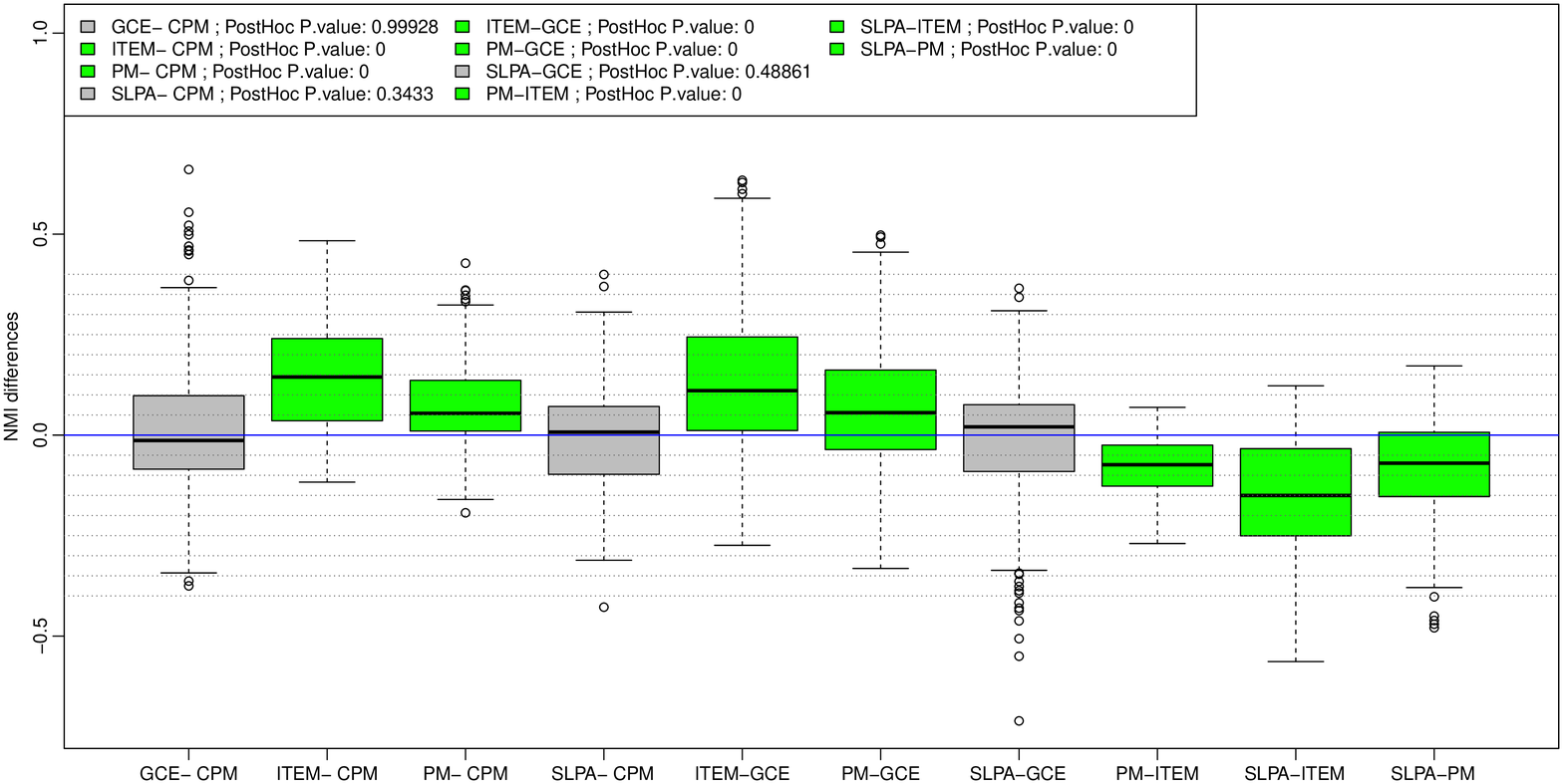}
 \includegraphics[width=65mm,height=50mm]{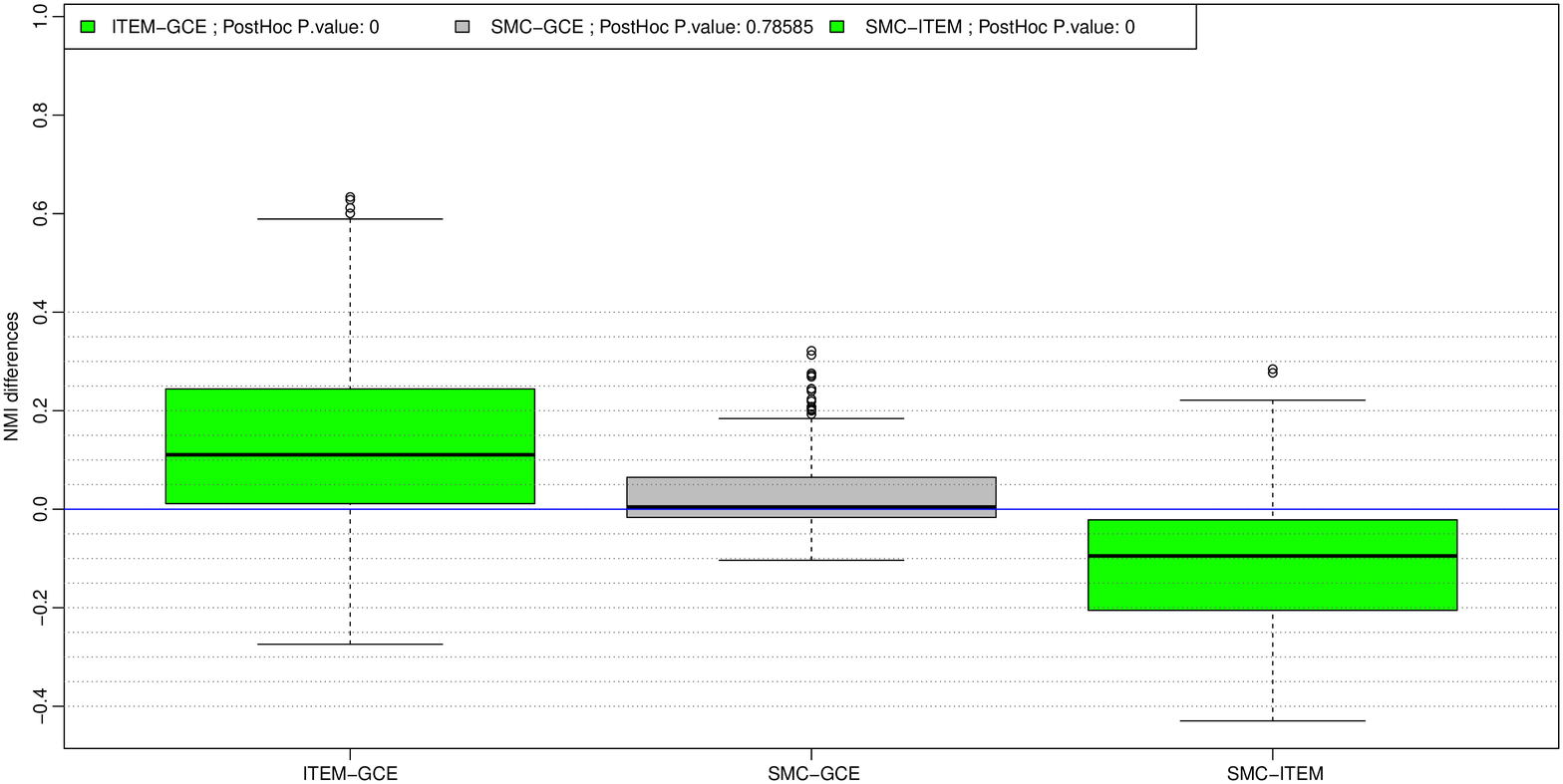}
 \end{tabular}   
 \caption{The PostHoc analyses of pairwise algorithms. The y axis indicates the NMI difference between two algorithms, and the green colored box indicates the two corresponding algorithms are significantly different (the significant level is 0.05). The p-value less than 1.0e-05 is rounded down to 0. The left subfigure compares ITEM, GCE, SLPA, CPM, and PM. The right subfigure compares ITEM, GCE, and SMC.}
\label{fig:statis}
\end{figure*}

\subsection{The comparing algorithms}

Many algorithms are devoted to discovering overlapping communities. Some algorithms \cite{gargi2011large,subbiancommunity,yang2009combining} use the domain or prior knowledge to promote final detection of communities. For example, \cite{gargi2011large} uses view counts of videos to find the most popular videos as seeds. But this information is not always available, and this paper selects algorithms that only use adjacent information of the detecting networks. We also tried to compare algorithms proposed by \cite{gopalan2013efficient,andersen2012overlapping,gleich2012vertex}, but these codes were difficult to deploy on our platform.

Xie et al. conduct a good survey of overlapping algorithms \cite{Xie2013overlapping} and indicate that GCE, Speaker-listener Label Propagation Algorithm (SLPA) \cite{xie2011slpa}, Community Overlap PRopagation Algorithm (COPRA) \cite{gregory2010finding}, Order Statistics Local Optimization Method (OSLOM) \cite{lancichinetti2011finding}, and LFM achieve higher performance on larger networks. This paper only compares ITEM with GCE and SLPA, while the other three algorithms are filtered out for their higher computational complexity compared with GCE and SLPA. Clique Percolation Method (CPM) \cite{palla2005uncovering} is also included for its high reputation. Lastly, we compare ITEM with the Poisson Model (PM) algorithm \cite{ball2011efficient}, which is considered a state-of-the-art method.

CPM first finds all $k$-cliques, then rolls one clique to another clique if they share $k-1$ nodes. The rolling process stops only when no adjacent cliques sharing $k-1$ nodes exist. All nodes covered by rolling $k$-cliques in the process form a community. GCE uses maximal cliques as seeds, then expands the nodes by maximizing the fitness function $f(C_i)=2 \times m_{in}^{C_{i}}/(2 \times m_{in}^{C_{i}}+m_{out}^{C_{i}})$, where $m_{in}^{C_{i}}$ is the number of edges in community $C_i$, and $m_{out}^{C_{i}}$ is the number of edges on the boundary of $C_i$. SLPA uses label propagation to discover overlapping communities, and nodes send and receive labels according to the sending and receiving rules. SLPA uses a threshold variable $r$ to filter out labels whose probability is lower than $r$. PM evaluates a set of parameters $\theta_{ik}$, which measure the extent of belonging for node $i$ to community $C_k$. 
 
Comparing ITEM with GCE, SLPA, CPM, and PM can show whether ITEM is superior to them, but cannot tell where the superiority comes from if it does exist. We design another algorithm to see the effects of our seeding and expanding method. As the expanding process of GCE is equal to minimizing the conductance between $C_i$ and the rest of the network, a mixing algorithm called Seeding and Minimizing Conductance (SMC) is built that uses the seeding process of ITEM with the expanding process of GCE. It is clear that SMC and GCE are identical except for their seeding methods, and SMC and ITEM are also identical except for their expanding methods. So from the performance difference between GCE and SMC, or the difference between SMC and ITEM, the effects of our seeding and expanding method can be concluded.

To run CPM, GCE, ITEM, and SMC, at least one parameter must be provided. For CPM and GCE, the clique level $k$ can range from 3 to 6, and the other parameters for GCE are set to their default values. The $th$ value for ITEM and SMC is in \{0.2, 0.3, 0.4\} for LFR networks; for Facebook100 networks, $th$ is set to 1.0. The $r$ value for SLPA is set in \{0.01, 0.05, 0.1, 0.15, 0.2, 0.25, 0.3, 0.35, 0.4, 0.45, 0.5\}. To run PM, the number of communities must be given. Hence, we only evaluate PM on LFR synthetic networks by providing the true number of communities to it. A threshold value $\theta_{th} \in \{ 0.6, 0.8, 1.0, 1.2, 1.4 \} $ is also used to filter out communities $C_k$ of $v_i$ if $\theta_{ik} < \theta_{th} $. For each network, all algorithms are run multiple times by iterating through their parameters and picking the best result.

\section{Experiments and analyses}

We conduct a wide range of experiments on the LFR networks. Figure \ref{fig:om} shows the performances of algorithms when changing the number of memberships of nodes. For each subfigure in figure \ref{fig:om}, there is a corresponding parameter setting generating it. We use double subscript notation to clearly cite a subfigure. A colon is used to cite all subfigures in a row or column. For example, fig\ref{fig:om}[2,3] is used to refer the subfigure at the $2$nd row and $3$rd column of figure \ref{fig:om}, fig\ref{fig:om}[2,:] refers to the four subfigures in the $2$nd row, and fig\ref{fig:om}[:,3] refers to the eight subfigures in the $3$rd column.

We first compare ITEM with GCE, SLPA, CPM, and PM. From figure \ref{fig:om}, it can be seen that no silver-bullet algorithm can surpass others over all 32 subfigures because of the diversity of the 32 parameter settings. But in most experiments, ITEM performs better than the other four algorithms. To see whether there is a statistical difference among ITEM, GCE, SLPA, CPM, and PM, we perform a Friedman test on an NMI result set that has 5 treatments (ITEM, GCE, SLPA, CPM, and PM) and 224 blocks (32 parameter settings and 7 options of $om$).  The p-value (2.2e-16) of the Friedman test shows that statistical differences exist among 5 algorithms. To further identify which algorithms are different, we carry out post hoc analyses using the paired Wilcoxon test; the results are shown in the left subfigure of figure \ref{fig:statis}. It can be seen that both ITEM and PM are significantly different from GCE, SLPA, and CPM. More specifically, ITEM can improve NMI more than 0.1 in about half of experiments; for PM, 0.05 improvement is achieved. The 32 subfigures in figure \ref{fig:om} also demonstrate the wide applicability of ITEM. Compared with other algorithms, ITEM always gives decent performances in more subfigures.

Then we compare ITEM with GCE and SMC. The same statistical analysis is carried out, except the original treatments are replaced with ITEM, GCE, and SMC. The results are shown in the right subfigure of figure \ref{fig:statis}. By comparing SMC with GCE, the effectiveness of our seeding method can be evaluated by comparing the method used by GCE, which is considered the best method \cite{lee2011seeding} for seeding. The right subfigure indicates that the qualities of seeds selected by SMC and GCE are comparable because there is no statistical difference between SMC and GCE. But the computational complexity of finding committees is lower than for finding all cliques. As committees are often dense sub-graphs that are similar to cliques, finding committees as seeds implies relaxing the seeding criteria but reaping the high efficiency. As for the effectiveness, we believe it should be attributed to the following reasons: first, the local ranking method of RSS overcomes the dilemma of the global ranking method. Second, MGIG selects seeds from a bird's eye perspective. Third, using committees as seeds is more proper than using cliques because clique criteria are too strict.  

 \begin{figure*}[!ht]
  \centering
\includegraphics[width=175mm,height=120mm]{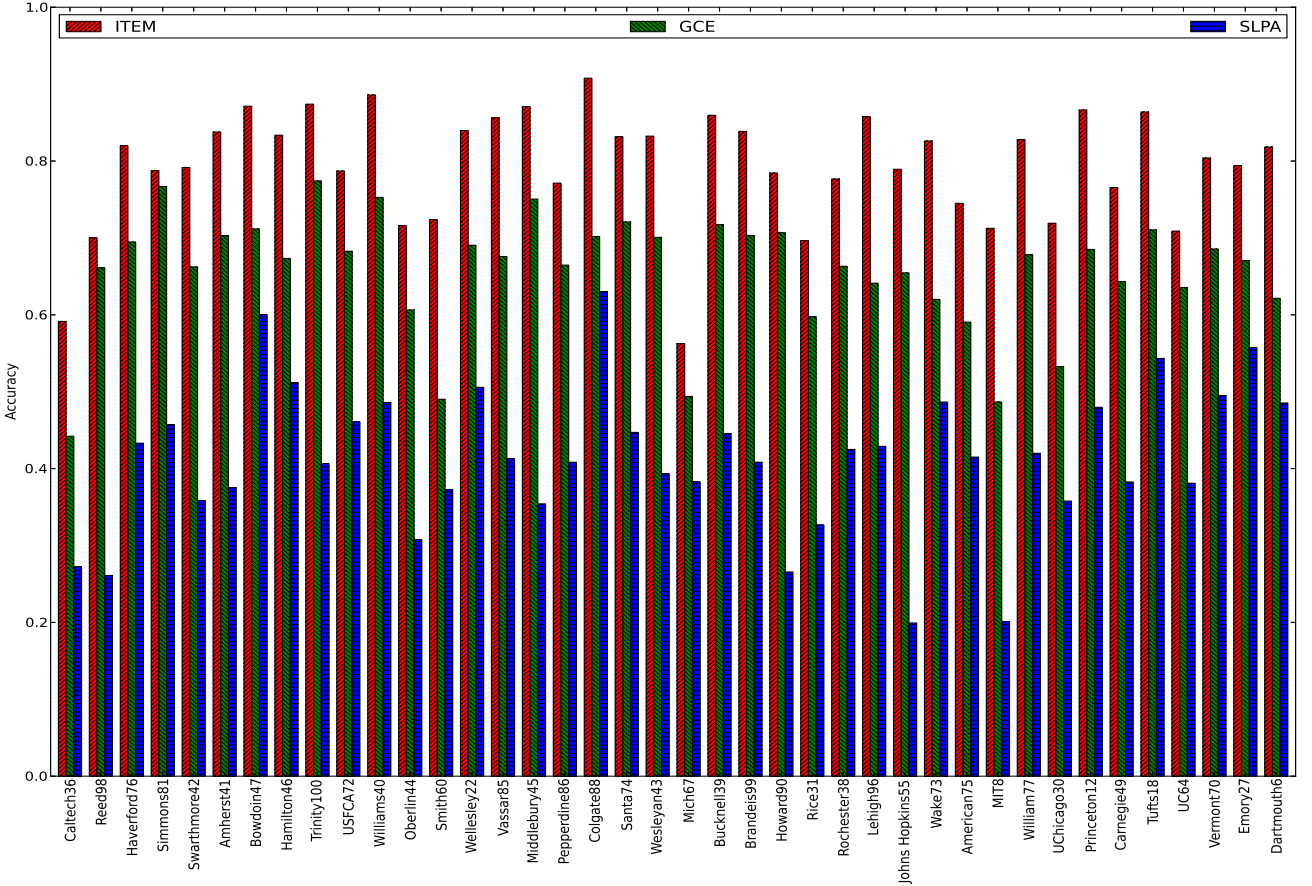}
 \caption{Accuracy comparisons among ITEM, GEC, and SLPA when using ``year'' as the label.}
\label{fig:year}
\end{figure*}

\begin{figure*}[!ht]
  \centering
\includegraphics[width=175mm,height=120mm]{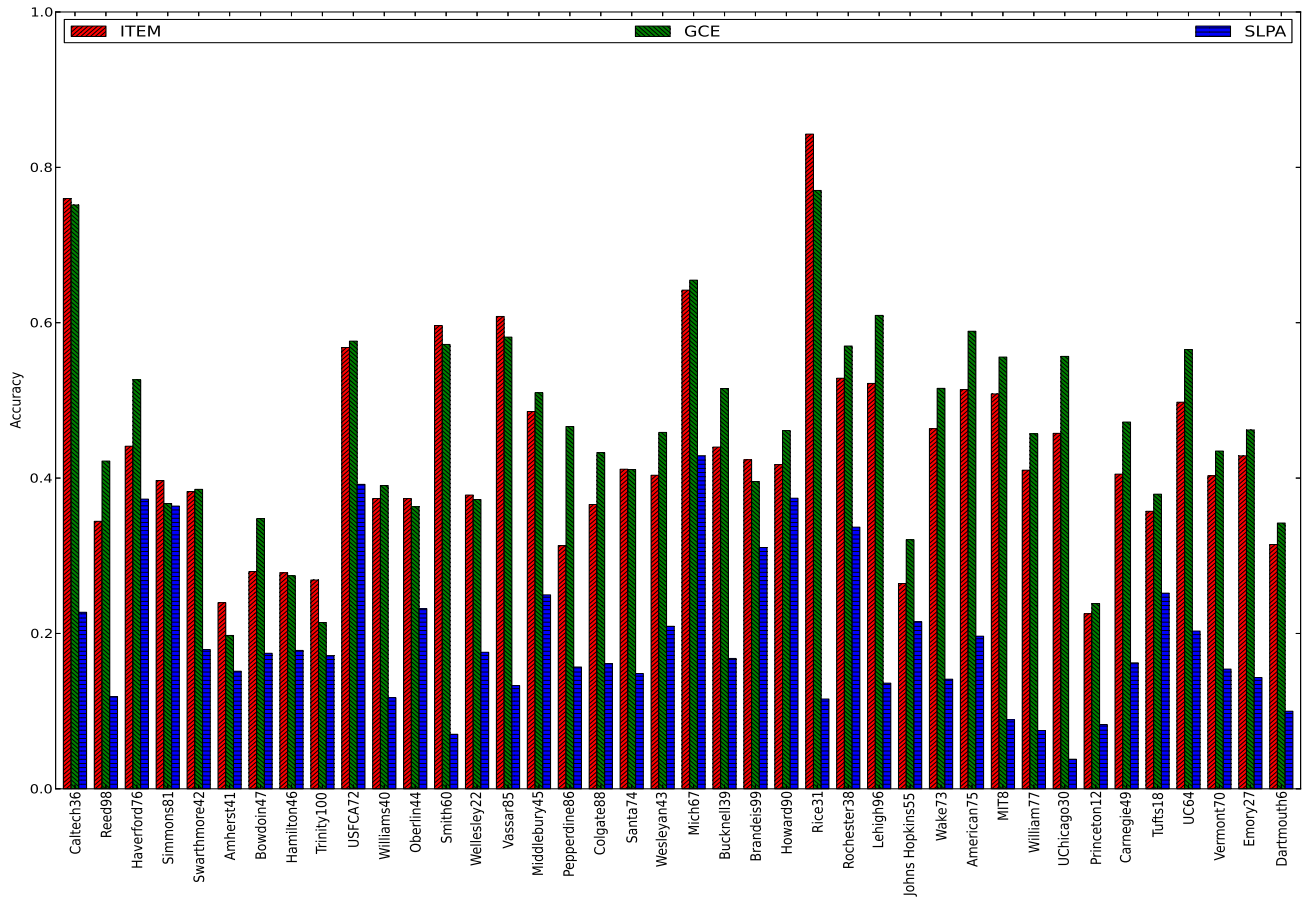}
 \caption{Accuracy comparisons among ITEM, GEC, and SLPA when using ``dorm'' as the label.}
\label{fig:dorm}
\end{figure*}

A significant difference exists between ITEM and SMC. The difference indicates that the expanding method using semi-supervised learning and Bayesian inference is superior to the heuristic method used by GCE. The following lists the differences between the former (i.e., ITEM's expanding method) and the latter (i.e., GCE's), and may explain where the superiority comes from: first, Bayesian inference gives the former a chance to promote the weights of authoritative nodes and suppress the weights of hub nodes. The heuristic functions used by the latter and other methods in \cite{yang2012defining} treat all nodes equally. The universality of Bayesian inference also renders the former wide applicability. Second, semi-supervised learning uses both labeled and unlabeled edges to model the NB classifier. Though unlabeled edges may theoretically decrease the performance of an NB classifier \cite{chapellerisks}, the high similar document presentation of edges in the same communities, plus the conservative strategy of only adding potential edges in each iteration, render an ideal scenario for applying semi-supervised learning. Third, by trying to minimize the global objective function (\ref{kmeansfunc}), competition is introduced by the former. In other words, each belonging is determined by the consultations of multiple communities, not only by a local fitness function.

We also conduct experiments using the Facebook100 dataset. Forty networks with the fewest nodes are used to evaluate communities found by ITEM, GCE, and SLPA. CPM is excluded because it cannot finish running in 48 hours. For each combination of network and algorithm, figure \ref{fig:year} and figure \ref{fig:dorm} display the accuracy given by the training classifier when using ``year'' and ``dorm'' attributes as labels, respectively. Each value of the bar is averaged over three times. Figure \ref{fig:year} indicates that ITEM is consistently better than GCE and SLPA on all 40 networks. For figure \ref{fig:dorm}, GCE performs better than ITEM and SLPA on most networks. We also probe the reasons why ITEM performs better than GCE when using ``year'' as a label but performs worse when using ``dorm'' as a label. This happens first because ITEM tends to generate larger found communities than GCE. Second, the true communities with ``year'' attributes are larger than the true communities with ``dorm'' attributes. When combining these two reasons, it can be seen that communities found by ITEM are more suitably used to infer the ``year'' attribute, while GCE is more suitable for the  ``dorm'' attribute. The second reason also explains why both ITEM and GCE get higher accuracy in figure \ref{fig:year} than in figure \ref{fig:dorm} for most networks. This is because the benchmark classifier has more training examples when using ``year'' to infer. The different behaviors of GCE and ITEM also enlighten us that combining multiple orthogonal algorithms may be a good choice to discover more true communities in a network.

It is also worth comparing ITEM with Linkcomm \cite{ahn2010link} because they both cluster edges into communities. But there is an important difference between ITEM and Linkcomm. we called ITEM a partitioning method because of the resemblance between K-means and our expanding method. Linkcomm uses the hierarchical agglomerative algorithm to cluster edges. As a consequence, it inherits two shortcomings of the hierarchical agglomerative algorithm. First, the greedy nature of agglomerating algorithms will yield sub-optimal clusters as compared with partitioning algorithms, because partitioning algorithms explore collective information to generate edge clusters, while agglomerating algorithms merely exploit two clusters' information at each agglomeration step \cite{ding2002cluster}. Second, one edge cannot change its belonging community once it is assigned to that community (most expanding methods also have this drawback). We believe these reasons can partly explain why Linkcomm gives poor performances \cite{Xie2013overlapping}.


\begin{table}[!ht]
\centering
\begin{tabular}{c|c|c|c|c}
\hline
Networks & m  & n  & K  & Running times (s)\\
\hline
Enron  & 367662 & 36692 & 2342 & 48.43 \\
Amazon & 925872 & 334863 & 25494 & 452.81 \\
Dblp  & 1049866 & 317080 & 30906 & 709.14 \\
\hline
\end{tabular}
\caption{The running times of ITEM on three empirical networks. The number of communities is found by ITEM.}
\label{tab:runtimes}
\end{table}

The computational complexity of ITEM's expanding method is $O(tpm)$, where $t$ is the number of iterations of the EM algorithm. The value of $t$ may be associated with $log(n)$ because the average path length is proportional to $log(n)$ for ``small-world'' networks \cite{newman2003structure}. In our experiments, the expanding process is stopped when only a few edges change labels. The overall computational complexity of ITEM is about $O(tpm)$ because its seeding complexity can be ignored in most cases when compared with its expanding complexity. Hence, ITEM is a proper choice to detect communities for large networks. Table \ref{tab:runtimes} lists the running times of ITEM on some empirical networks \footnote{http://snap.stanford.edu/data/index.html}. All experiments are conducted on a workstation that has 8 Intel Xeon 2.27GHz cores and 12G RAM. ITEM is paralleled running on 8 cores using openMP. It can be seen that ITEM can process Dblp network in 12 minutes. By virtue of the ideal parallel parallelism of K-means \cite{stoffel1999parallel}, we believe ITEM can process even larger networks with more machines efficiently.

\section{Conclusion and future works}
In this work, an overlapping community detection algorithm called ITEM has been presented. ITEM is devoted to solve the problems of how to efficiently select high-quality seeds and to make the expanding method applicable to a wide range of networks. To solve the first problem, seeds are selected using local and global methods. For the second problem, we resort to the semi-supervised learning and NB classifier. The experimental results show the advantage of our seeding and expanding methods. The statistical analysis further demonstrates that ITEM improves performance significantly when compared with most existing algorithms.

ITEM can only run on unweighted and undirected networks now. In the future, we will extend it on weighted and directed networks. ITEM may be treated as a preparatory step for other tasks, such as link prediction and key point detection, because after ITEM finishes we can get parameters such as $p(v_j|C_k)$ and $p(C_k|v_j)$. Lastly, it is also worth making ITEM available to incremental or dynamic networks. 

\section{Acknowledgments}
The authors would like to thank the anonymous referees for their comments. This research is partially supported by grants from the National Science Foundation of China (No.61303167 and No.61363047) and by funds from the Basic Research Program of Shenzhen (Grant No. JCYJ20130401170306838).

\bibliography{item}  

\end{document}